\documentclass[12pt]{article}
\usepackage{graphicx}
\begin{document}

\pagestyle{empty}

\noindent
{\bf A Study of Multiple-Mode Pulsating Red Giants}

\bigskip

\noindent
{\bf John R. Percy and Tanya Brekelmans\\Department of Astronomy and Astrophysics\\University of Toronto\\Toronto ON\\Canada M5S 3H4}

\bigskip

{\bf Abstract}  This project began as a search for triple-mode pulsating
red giants (PRGs) among known double-mode PRGs, using AAVSO visual data.  
It then expanded to include a check on the reality and significance of some
double-mode PRGs.  We identified several ``double mode" PRGs
in which one of the periods might simply be an alias of the other.  We also comment
on the possibility that the additional periods may be harmonics, rather than
overtones.  As for triple-mode pulsators: in our sample, we find no convincing cases in
which there is a significant, non-alias third period which might be an
additional radial period.  We note, however, that a few triple-mode pulsators have been
found in a very large sample of PRGs in the LMC.  We discuss the results in the context
of recent theoretical models of pulsating red giants.

\medskip

\noindent
{\bf 1. Introduction}

\smallskip

Red giants are unstable to pulsation.  The period and amplitude both increase
as the star expands, brightens, and cools.  Initially, the pulsation is driven
by the turbulent pressure of convection in the outer layer of the star, just as solar
oscillations are.  As the star expands and cools, the standard opacity
mechanism also comes into play.

In a recent paper, Xiong and Deng (2007) report on new and improved theoretical models of
PRGs, and non-adiabatic pulsation analysis using a non-local time-dependent
theory of convection.  They find that, (i) at low luminosities, higher-order
radial modes are unstable but, (ii) as the luminosity increases, the fundamental
and first overtone modes become unstable, and the higher-order modes become
stable.  Finding (i) is consistent with the results of Percy and Polano (1998),
Percy and Parkes (1998), and Percy and Bakos (2003), who found that in
small-amplitude, short-period PRGs, higher-order modes were
excited: mostly the first overtone, occasionally the 2nd or 3rd or the
fundamental.  Finding (ii) is supported by the large number of more luminous PRGs which
are observed to be pulsating with two periods which are consistent with
the fundamental and first overtone.  Xiong and Deng (2007) also present theoretical periods and period ratios of low-order modes for
models of various masses and luminosities.

Perhaps the most fruitful study of multiple pulsation modes in PRGs is that
of large surveys in the Magellanic Clouds, e.g. Soszynski et al. (2013).  They considered a large (91,995!) sample of PRGs in the Large Magellanic Cloud
which, in period-luminosity diagrams, occupied a series of sequences. Among
other things: they find, among the brighter stars, a small group of about 50 triple-mode variables, with the fundammental, first overtone, and second overtone modes simultaneously excited.

Several observational studies using AAVSO visual observations have
identified luminous PRGs with two or three periods, usually the fundamental and
first overtone radial mode and, in some stars, a ``long secondary period" of
unknown cause (Nicholls et al. 2009).  The initial objective of our study was to analyze stars with
the fundamental and first overtone modes excited, to see if any of them also showed
the second overtone mode.  As we did this, we noticed that some of the
secondary periods published in previous studies were possibly one-cycle-per-year
aliases of the primary period.  A second objective of
our study was therefore to flag these possible alias periods.

\medskip

\noindent
{\bf 2. Data and Analysis}

\smallskip

We examined PRGs which had been shown, in the following studies,
to have two periods with a ratio close to 2.0, the expected ratio of the
fundamental and first overtone period: Mattei et al. (1997), Kiss et al. (1999),
Percy and Tan (2013), and Percy and Kojar (2013).
We used visual observations, from the AAVSO International Database.
Percy and Abachi (2013) discussed some of the limitations of visual
data which must be kept in mind when analyzing the observations, and
interpreting the results.
The data
were analyzed using the
{\it VSTAR} package (Benn 2013; www.aavso.org/vstar-overview), primarily the Fourier 
(DCDFT) analysis routine.  As noted by 
Kiss et al. (2006), the peaks in the power spectra of PRGs are complex,
rather than sharp, and it is not always possible to identify an exact period.
Furthermore: especially for the small-amplitude stars, the noise level and
complexity are such
that it is generally difficult to identify secondary periods.

\medskip

\noindent
{\bf 3. Results and Discussion}

\smallskip

\noindent
{\bf 3.1 Reality of Some Double-Mode Pulsating Red giants}

\smallskip

The first question is whether some of the secondary-mode peaks in the DCDFT
spectrum might actually be ``aliases" of the primary-mode peak.  Aliases
occur because of the seasonal gaps in the data, and their frequencies
are separated by $\pm$ N cycles per year from the true frequency, 
where N is most commonly 1.  In Table 1, we have listed the stars
for which, according to our analysis, the secondary-mode peak might be an alias,
rather that a second mode of pulsation.  The sources in the last column
are: (1) Kiss et al. (1999); (2) Mattei et al. (1997); (3) Percy and Tan (2013);
(4) Percy and Kojar (2013).  It is possible, of course, that a few stars
have a real period which is nearly coincident with the alias value.  As
Kiss et al. (1999) pointed out, the peaks in the power spectra of 
PRGs tend to be complex.

The second question arises from the fact that many apparent double-mode
PRGs have periods which differ by a factor of close to 
2.0: could the shorter period actually be a subharmonic of the longer
period, and due to the non-sinusoidal nature of the light curve?  For example: DCDFT
analysis of Delta Cephei, which has a non-sinusoidal light curve, shows a weak signal at twice the
fundamental frequency, and a very weak signal at three times the
fundamental frequency.  There
are several arguments against this possibility, for our PRGs: (1) the ratio of fundamental to first
overtone period is known, from theoretical models, to be close to 2, and longer-period 
PRGs are predicted to be unstable to the fundamental and first
overtone modes (Xiong and Deng 2007); (2) the actual ratio of the two
periods is not always exactly 2, but is observed and predicted to be a weak function of the longer period
(Kiss  et al. 1999; Percy and Tan 2013, Xiong and Deng (2007)); (3) if the shorter period
was the first subharmonic, we might expect to see a third peak at a period
of P/3 where P is the primary period, and this has not been reported; (4) the light curves and phase
curves are not noticeably
non-sinusoidal.  

We examined the following PRGs which had {\it strong} fundamental
periods, and a second period which is close to half of the first, to see whether
there was a third period which was about a third of the first: RV And, ST And,
TV And, V Boo, SV Cas, T Col, RZ Cyg, T Eri, and S Tri.  None of these
stars has a period near one-third of the fundamental period which is
not an alias of the fundamental or first overtone periods.

\begin{figure}
\begin{center}
\includegraphics[height=7cm]{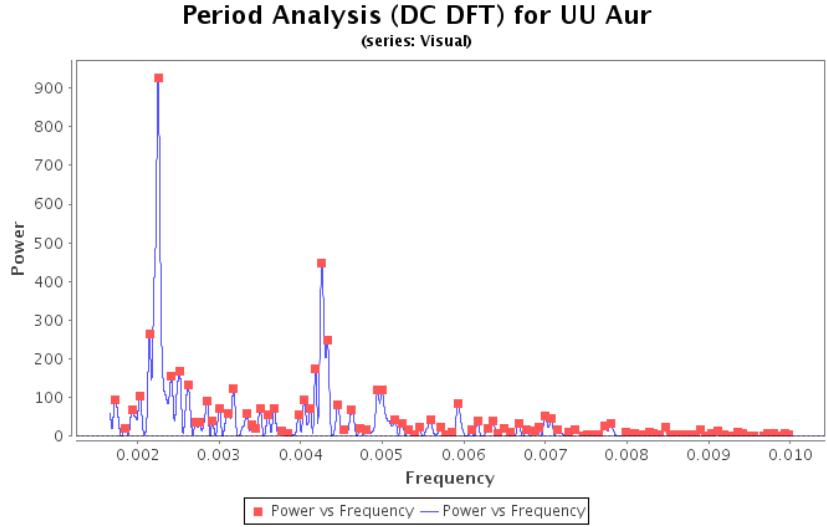}
\end{center}
\caption{The DCDFT spectrum of UU Aur.  The peaks at 0.0023 and 0.0043 cycles
per day are the fundamental and first overtone frequencies.  The peak at
0.005 is an alias.  The peak at 0.0059 cycles per day (period of
168.5 days) may be the second overtone frequency.}
\end{figure}

\medskip

\noindent
{\bf 3.2 Triple-Mode Pulsating Red Giants?}

\smallskip

We have examined the DCDFT spectra of several dozen double-mode pulsating red giants
in sources 1-4, to look for third periods which are not aliases of the
first or second.  Assuming that the third period is the second overtone,
the third period should be about 0.25-0.40 of the fundamendal period, according
to the models of Xiong and Deng (2007).  The best case that we can find,
and it is most likely spurious, is UU Aur (Figure 1).  This result is not
inconsistent with that of Soszynski et al. (2013), who examined a much larger
(over 90,000 stars) sample, and were using photometric rather than visual data.
It is also possible that there is a difference in behavior of LMC and
Galactic PRGs, due to the difference in metallicity.

\medskip

\noindent
{\bf 3.2A.  Notes on Individual Stars}

\smallskip

{\it UU Aur:} There is a small peak at a period of 168.5 days which could
possibly be a third period (Figure 1).  This is the best candidate for a triple-mode
PRG, though the likelihood is small.

\smallskip

{\it ST Her:} The 152- and 256-day periods appear to be aliases; the 372-day
period may be real, or it may be a spurious period caused by the Ceraski
effect, which is a physiological effect of the visual observing process,
which can cause apparent variability with a period of one year.  If it is real, it is difficult to interpret in terms of a radial
mode.

\smallskip

{\it RV Mon:} The 585- and 979-day periods are aliases.  There are peaks at
366$\pm$5 days which may be spurious, and due to the Ceraski effect.

\smallskip

{\it BQ Ori:} The periods other than 125 and 237 days appear to be aliases,
with the exception of 352 days which could possibly be spurious, and caused by
the Ceraski effect.

\smallskip

{\it RZ UMa:} The peaks are complex, but it appears that the 254- and 146.2-day
periods are aliases.

\medskip

\noindent
{\bf 3.3 Comparison with Theoretical Models}

\medskip

Xiong and Deng (2007), using theoretical models, have noted the presence
of a red giant instability strip, which is well-known from a century or
more of observations.  This strip extends {\it smoothly} from low-luminosity,
short-period stars to high-luminosity, long-period (Mira) stars.  This, too,
is well-known from observations, despite the tendency to subclassify the PRGs in various discrete
ways; see also Soszynski et al. (2013).

In the lower-luminosity stars, turbulent pressure due to convection is the
most important driving mechanism.  Percy and Abachi (2013) have discussed
the possibility that the ubiquitous variations in PRG pulsation amplitudes
are a result of this mechanism.  In higher-luminosity stars, the standard
kappa mechanism also contributes to the instability of the star.

The models also predict that, at lower luminosities, higher radial modes
are excited, whereas in higher-luminosity stars, the fundamental and first
overtone modes are excited, and higher modes are stable.  As noted in the
Introduction, and in sources 1-4, this is consistent with observations.

Xiong and Deng (2007) predict period ratios for low-order radial modes.
As noted by Kiss et al. (1999) and Percy and Tan (2013), the observed ratio of the
first-overtone period to the fundamental period is not constant, but varies
from about 0.56 for fundamental periods of 100 days, to about 0.50 for
fundamental periods of 500 days.  This is consistent with the {\it trend}
found by Xiong and Deng (2007), though the latter value (0.50) is somewhat
higher than the theoretical values found for masses between 1.0  and 3.0 solar.

If the 168.5-day period in UU Aur is real, then its ratio to the fundamental
period 0.263 is consistent with the models of Xiong and Deng (2007) for
any reasonable mass.

Finally: Kiss et al. (1999) and Mattei et al. (1997) list 61 PRGs which have two
periods with a ratio close to 2.0, and which are therefore likely to be the
fundamental and first overtone periods.  They also list the amplitude of
each.  We note that, in almost every case, the amplitude of the fundamental
period is greater than that of the first overtone period.  There are very
few cases in which the first overtone is the dominant mode.  We also note
that the ratio of the first-overtone amplitude to the fundamental amplitude
does {\it not} seem to depend on the fundamental period.

\medskip

\noindent
{\bf 4. Conclusions}

\smallskip

Although Soszynski et al. (2013) have found a very small fraction of a large
sample of PRGs in the LMC to be triple-mode pulsators, we have not found
any such stars in a sample of several dozen luminous PRGs for which we have systematic,
sustained visual observations in the AAVSO International Database.

\medskip

\noindent
{\bf Acknowledgements}

\smallskip

We thank the hundreds of AAVSO observers who made the observations which were
used in this project, the AAVSO staff who processed and
archived the measurements, and the team which developed the
{\it VSTAR} package and made it user-friendly and publicly available.
Coauthor JRP thanks coauthor TB, an undergraduate astronomy and physics
major, who carried out most of this project for her ``senior thesis".
This project
made use of the SIMBAD database, which is operated by CDS,
Strasbourg, France.

\medskip

\noindent
{\bf References}

\smallskip

\noindent
Benn, D. 2013, {\it VSTAR} data analysis software (http://www.aavso.org/node/803).

\smallskip

\noindent
Kiss, L.L. et al., 1999, {\it Astron. Astrophys.}, {\bf 346}, 542.

\smallskip

\noindent
Mattei, J.A. et al., 1997, in {\it Hipparcos Venice 1997}, ESA SP-402, 269.

\smallskip

\noindent
Nicholls, C.P., Wood, P.R., Cioni, M.-R.L., and Soszynski, I., 2009,
{\it Mon. Not. Roy. Astron. Soc.}, {\bf 399}, 2063.

\smallskip

\noindent
Percy, J.R. and Parkes, M., 1998, {\it Publ. Astron. Soc. Pacific}, {\bf 110}, 1431.

\smallskip

\noindent
Percy, J.R. and Polano, S., 1998, in {\it A Half-Century of Stellar Pulsation
Interpretations}, ed. P.A. Bradley and J.A. Guzik, ASP Conference Series, \#135,
249.

\smallskip

\noindent
Percy, J.R. and Bakos, A.G., 2003, in {\it The Garrison Festschrift}, ed.
R.O. Gray et al., L. Davis Press, 49.

\smallskip

\noindent
Percy, J.R., and Tan, P. 2013, {\it JAAVSO}, {\bf 41}, 1.

\smallskip

\noindent
Percy, J.R., and Kojar, T., 2013, {\it JAAVSO}, {\bf 41}, 15.

\smallskip

\noindent
Percy, J.R. and Abachi, R., 2013, {\it JAAVSO}, {\bf 41}, 193. 

\smallskip

\noindent
Soszynski, I., Wood, P.R., and Udalski, A., 2013, {\it Astrophys. J.}, {\bf 779}, 167, pp. 6.

\smallskip

\noindent
Xiong, D.R., and Deng, L., 2007, {\it Mon. Not. Roy. Astron. Soc.}, {\bf 378}, 1270.

\bigskip

\small

\begin{center}
\begin{table}
\caption{Possible Alias Periods in Pulsating Red Giants.} 
\begin{tabular}{rrrr}
\hline
Star & P1(d)/$\Delta$mag & P2(d)/$\Delta$mag & Source \\
\hline
RV And & 166.4/0.76 & 88.8/0.13 & 2 \\
V Aqr & 689/0.25 & 242/0.35 & 1 \\
Y CVn & 273/0.06 & 160/0.05 & 1 \\
T Cet & 299.5/0.34 & 162.4/0.45 & 2 \\
TT Cyg & 390/0.03 & 188/0.03 & 1 \\
U Del & 1146/0.21 & 580/0.05 & 1 \\
RV Mon & 585/0.1 & 979/0.15 & 3 \\
RU Per & 328.7/0.38 & 170.4/0.42 & 2 \\
RZ UMa & 147/0.03 & 247/0.05 & 4 \\
R UMi & 325/0.42 & 170/0.11 & 1 \\
\hline
\end{tabular}
\end{table}
\end{center}

\end{document}